\documentclass[preprint,12pt]{elsarticle}

\usepackage{lineno}

\usepackage{epsfig}
\usepackage{array,tabularx,epsfig,mathrsfs,graphicx,rotating}
\usepackage{amsmath}
\usepackage{ifthen}
\usepackage{comment}
\usepackage{ragged2e}
\PassOptionsToPackage{hyphens}{url}
\usepackage[hyphens]{url}
\usepackage{hyperref}
\usepackage{listings}
\usepackage{epstopdf}
\usepackage{color}
\usepackage{float}
\usepackage{subfig}
\usepackage{subcaption}

\hypersetup{
  colorlinks=true,
  linkcolor=blue,
  citecolor=blue,
  urlcolor=blue
}

\graphicspath{{figs/}}

\journal{xxx}

\begin{document}

\begin{frontmatter}

\title{Design and experimental application of a radon diffusion chamber for determining diffusion coefficients in membrane materials}

\author[a]{L.~Wu}

\author[a]{L.~Si}

\author[a]{Y.~Wu}

\author[a]{Z.~Gao}

\author[b]{Y.~Heng}

\author[a]{Y.~Li}

\author[c,a,d]{J.~Liu}

\author[b]{X.~Luo}

\author[e]{F.~Ma}

\author[a,d]{Y.~Meng\corref{cor}%
}
\ead{mengyue@sjtu.edu.cn}

\author[b]{X.~Qian}

\author[a]{Z.~Qian}

\author[a]{H.~Wang}

\author[a]{Y.~Yun}

\author[f]{G.~Zhang}

\author[b]{J.~Zhao}

\address[a]{School of Physics and Astronomy, Shanghai Jiao Tong University,\\
Key Laboratory for Particle Astrophysics 
and Cosmology (MoE), 
\\ Shanghai Key Laboratory for Particle Physics and Cosmology, Shanghai 200240, China}

\address[b]{Institute of High Energy Physics, Chinese Academy of Sciences, Beijing, 100049, China}

\address[c]{New Cornerstone Science Laboratory, Tsung-Dao Lee Institute, Shanghai Jiao Tong University, Shanghai, 200240, China}

\address[d]{Shanghai Jiao Tong University Sichuan Research Institute, Chengdu 610213, China}

\address[e]{Vanpek Enterprises Beijing Inc, Beijing, 102200, China}

\address[f]{Jiangsu Donchamp New Material Technology Co., Ltd, Taixing City, 225400, China}

\cortext[cor]{Corresponding author}

\begin{abstract}

In recent years, the issue of radon emanation and diffusion has become a critical concern for rare decay experiments, such as JUNO and PandaX-4T. This paper introduces a detector design featuring a symmetric radon detector cavity for the quantitative assessment of membrane materials' radon blocking capabilities. The performance of this design is evaluated through the application of Fick's Law and the diffusion equation considering material solubility. Our detector has completed measurements of radon diffusion coefficients for four types of membrane materials currently used in experiments, which also confirms the rationality of this detector design. The findings are instrumental in guiding the selection and evaluation of optimal materials for radon shielding to reduce radon background, contributing to boost sensitivities of rare event research.
\end{abstract}


\end{frontmatter}


\section{Introduction}

Rare event detection experiments have extremely stringent requirements for background control, and these experiments strive to maintain a minimal background level to reduce interference with the desired signal. In practice, there are various sources of experiment backgrounds, including radioactive elements in the construction materials of the detectors, radioactive particles originating from cosmic rays, and even the surrounding environment of the detectors\cite{PandaX-4T:2021lbm, XENON:2020fbs}. In many of today's experiments, radon has been a notable contributor to the background. The beta and gamma signals emitted in the decay chain of radon and the neutrons released by its decay products serve as a prominent background signal in the analysis of many physics topics. Therefore, preventing radon from entering sensitive areas of experiments is crucial.

In practice, people will use membrane materials to block radon and its decay daughters. In some experiments, such as the Jiangmen Underground Neutrino Observatory (JUNO)\cite{2022103927,JUNO:2021kxb}, a coating is applied to the surface of the detector during the construction process to block radon gas. Additionally, some membrane materials are used to protect the cleaned detector components during transportation to prevent radon in the air from entering the components. However, the insulating effect of materials is inherently limited due to the diffusion of radon gas. In evaluating this performance, people have designed several devices to measure the diffusion coefficient of materials, ranging from membrane materials\cite{DAOUD2001127, JIRANEK20091318} to others such as concrete and air\cite{TSAPALOV20147}. In this paper, we developed a novel detector that comprises a symmetric radon detector cavity, building upon the method of electrostatic collection of radon daughter products. The use of the silicon PIN diode enhances the radon detection efficiency and is offering greater accuracy. This new compact detector can better determine the radon diffusion coefficient of various membrane materials while ensuring the measurement system remains sufficiently flexible and controllable.

\section{Radon Diffusion Model}

The physical process of radon diffusion is typically characterized by Fick's Law and the diffusion equation:
\begin{equation}\label{eq1}
    J=-D\nabla C
\end{equation}
\begin{equation}\label{eq2}
    D\nabla^2C-\lambda C=\frac{\partial C}{\partial t}
\end{equation}

where $J$ is the rate of radon flux per unit area, $D$ is the diffusion coefficient, $C$ is the radon number concentration in the chamber, and $\lambda$ is the radon decay constant.

\begin{figure}[!ht]
	\centering
	\includegraphics[width=0.7\linewidth]{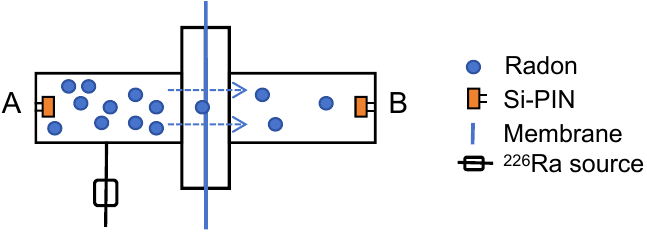}
 \caption{Schematic diagram of radon diffusion within the detector chamber.}
	\label{fig:diffuse_process}
\end{figure}

The illustration of the radon diffusion process within our detector is shown in Fig.\ref{fig:diffuse_process}. In such studies, it is essential to take into account the solubility of the material, as it plays a significant role in the relationship between radon concentrations inside and outside the material. The consideration of solubility will be detailed later. However, when the solubility is not well understood, we can first calculate the average diffusion coefficient to gain insight into the material’s overall effectiveness in blocking radon. We do not need to consider the solubility when we aim to solve for the average diffusion coefficient. By focusing on the radon number concentration $C_B$, with B representing the side opposite the source, we can derive the following equation based on equations (\ref{eq1}) and (\ref{eq2}) by assuming a one-dimensional space for simplification:
\begin{equation}\label{eq4}
    \frac{\partial C_B}{\partial t}=-\frac{D_a\Omega}{V_B}\cdot\frac{C_B-C_A}{\tau}-\lambda C_B
\end{equation}

where $D_a$ is the average diffusion coefficient, $\Omega$ is the area of membrane material, $C_A$ is the radon number concentration on the A side, $\tau$ is the thickness of the membrane, and $V_B$ is the volume of the chamber on the B side. When the system reaches the equilibrium, $C_B$ remains constant, which leads to the left side of the equation becoming zero. Then we can rewrite equation (\ref{eq4}) in an equilibrium state as:
\begin{equation}\label{eq5}
    D_a=\frac{\tau V_B\lambda }{\Omega}\cdot\frac{\eta}{1-\eta}
\end{equation}

Here, $\eta$ represents the ratio of radon number concentration on either side of the membrane:
\begin{equation}\label{eq6}
    \eta=\frac{C_B }{C_A}
\end{equation}

Therefore, once the concentration ratio across the membrane is known, we can readily determine the average diffusion coefficient $D_a$, which characterizes the macroscopic effect of the membrane material on radon diffusion.

When examining radon solubility $S$ in materials, we can posit a proportional correlation between radon levels inside and outside the membrane surface, as indicated by equation (\ref{eq3}). This conclusion has been mentioned in prior research\cite{WOJCIK19918}.
\begin{equation}\label{eq3}
    C_i' = SC_i
\end{equation}

Here, $C_i$ is radon concentration outside the membrane, $C_i'$ is the concentration inside, and $S$ indicates the solubility of radon in the membrane relative to the gas phase. At time $t$ in an equilibrium state, we can get $C_i(t)$ by the radon detectors distributed on both sides of the membrane. The following sections will outline the procedure for acquiring these parameters by assessing the radioactivity of radon daughters on both sides of the membrane using our system.

To delve deeper into the relationship between the exact diffusion coefficient and solubility, we can formulate an expression similar to that in equation (\ref{eq4}):
\begin{equation}\label{eq7}
    \frac{\partial C_B}{\partial t}=-\frac{D\Omega}{V_B}\frac{d C_B'}{dx}-\lambda C_B
\end{equation}

Once again, when the system reaches an equilibrium state, the left side of the equation (\ref{eq7}) becomes 0. We adopt the processing method outlined in article\cite{MENG2020108963} to obtain the same transcendental equation:
\begin{equation}\label{eq8}
\alpha\cdot\eta=\frac{S\cdot \Omega}{V_B}\left[\frac{1-\eta\cdot cosh(\alpha\cdot \tau)}{sinh(\alpha\cdot \tau)} \right]
\end{equation}

where we have followed the convention in the context:
\begin{equation}\label{eq9}
 \alpha=\sqrt{\frac{\lambda}{D}}
\end{equation}

Currently, the exact value of $S$ for the material cannot be determined with our experimental design. To address this, we vary the solubility parameter $S$ between 1 and 20 in our calculations for each tested material. The ratio of radon number concentrations on both sides of the membrane can be measured experimentally, so the value of $D$ in the equation (\ref{eq8}) can be solved by combining the given value of $S$, which will lead to a diffusion coefficient as a function of solubility.
\section{Methodology}\label{sec:artwork}
\subsection{Experimental Setup}

\begin{figure}[!b]
	\centering
\includegraphics[width=0.7\linewidth]{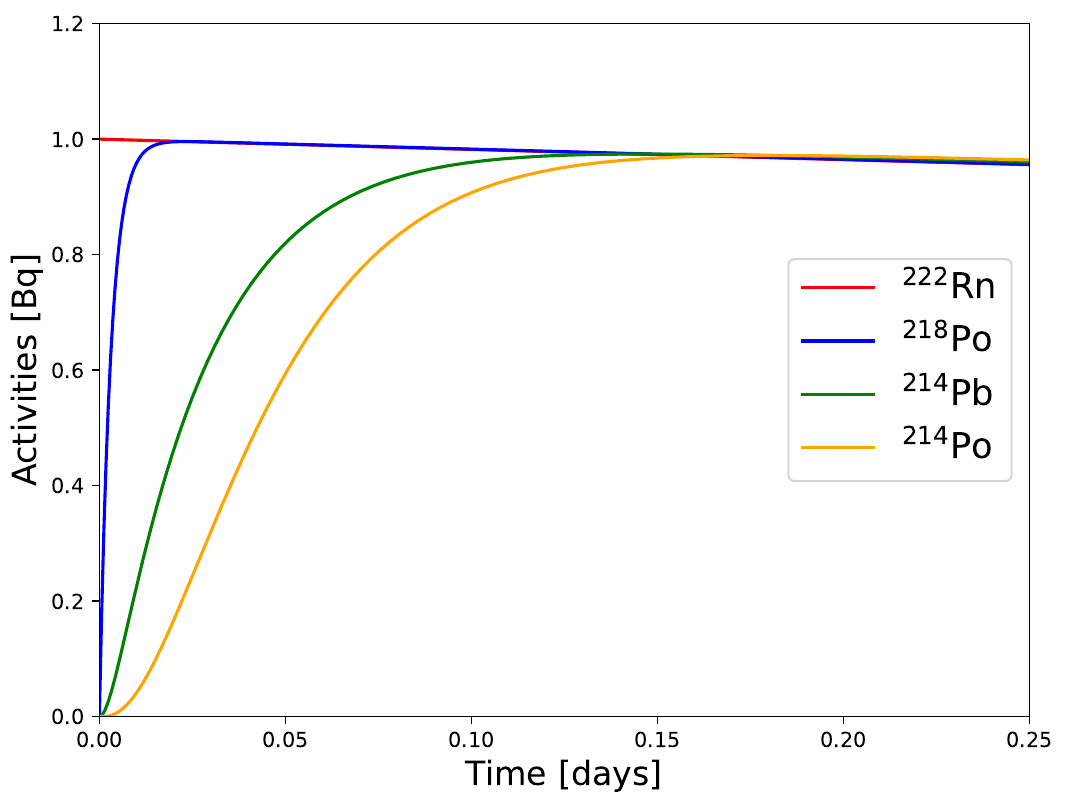}
 \caption{The variation in radioactive activity of different isotopes over time. The scenario assumes an initial introduction of 1 Bq of ${}^{222}$Rn into the chamber. It shows that approximately after 0.17 days, the radioactive activities of various elements tend to converge.}
	\label{fig:pic1}
\end{figure}

The classical method of radon measurements, known as electrostatic collection of radon daughters, primarily consists of three components: the collection cavity, signal processing section, and signal acquisition section. More details about this method can be found in reference\cite{Mott:2013nvz}. This section will primarily introduce the setup of our experiment.

The collection cavity effectively gathers the charged daughters of radon, with a primary focus on ${}^{214}$Po and ${}^{218}$Po, onto the silicon PIN diode, utilizing a well-established and stable electrostatic field. Since most of the radon's decay products are positively charged (about 87\%)\cite{PhysRevC.92.045504}, the electrostatic field will drag these positive particles to the surface of the diode. The energy emitted during the alpha decay of these radon daughters will deposit within the silicon PIN diode, and then we can collect these electrical signals. After the decay chain reaches equilibrium, as shown in Fig.\ref{fig:pic1}, the radioactive activity of radon and daughters like polonium will also reach equilibrium. We can measure the radioactive activities of these polonium daughters to obtain the radon number concentration in the cavity\cite{Mott:2013nvz}.

\begin{figure}[!h]
	\centering
\includegraphics[width=0.9\linewidth]{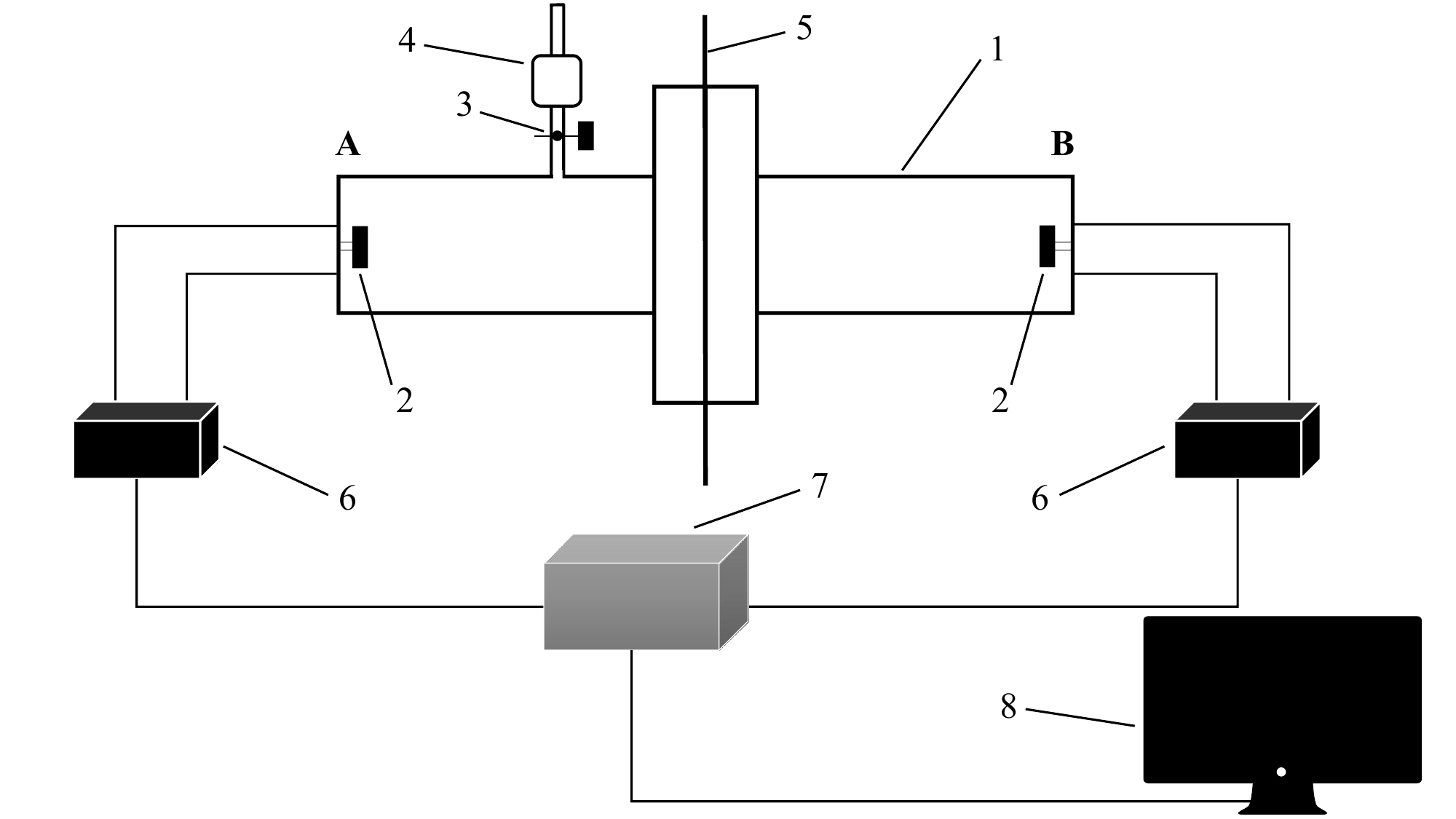}
 \caption{Diagram depicting the experimental arrangement: 1 - Stainless steel chamber; 2 - Silicon PIN diode; 3 - Ball valve; 4 - Radium source (${}^{226}$Ra\cite{SQIU200610,DZGJ200802015}); 5 - Membrane material for testing; 6 - Pre-signal amplifier box; 7 - Digital multi-channel analyzer (Digital MCA, DT5780\cite{DT5780MCA}); 8 - Desktop computer.}
	\label{fig:pic2}
\end{figure}

\begin{figure}[ht]
	\centering
\includegraphics[width=0.7\linewidth]{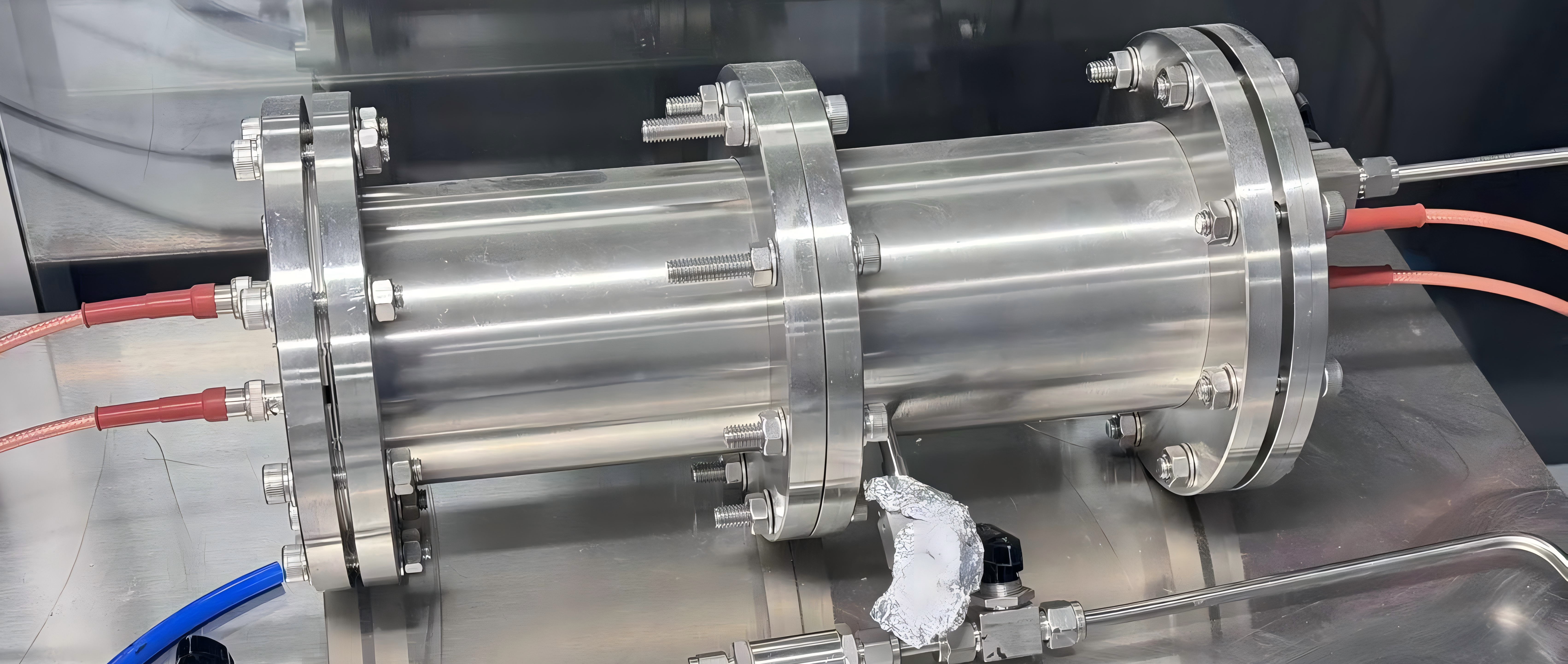}
 \caption{Physical image of the symmetric stainless steel radon detector chamber.}
	\label{fig:chamber_realpic}
\end{figure}

Fig.$\ref{fig:pic2}$ illustrates the experimental setup used in this study. The primary component consists of two symmetrical radon detector cavities with a diameter of 0.165 m (see Fig.\ref{fig:chamber_realpic}), connected by O-ring seals at the center. The test membrane material can be seamlessly inserted into the central section. Two silicon PIN diodes\cite{Si-PIN} are positioned at the end of each chamber to detect the alpha signals, and both of these PIN diodes will be supplied with negative high voltage. One of the cavities is linked to a radium source (${}^{226}$Ra), featuring a spherical valve positioned at its center to regulate the access to or containment of the radioactive source.

The silicon PIN diode generates an initial electrical signal, which undergoes amplification through a pre-signal amplifier before being sent to the multi-channel analyzer (MCA) and the personal computer (PC). We employ the \textit{COMPASS} (CAEN Multi-PArameter Spectroscopy Software) for real-time data readout and storage on the PC side. This software lets us view real-time energy spectra and capture spectral information over time. The evolution of the radon decay chain within the chamber is encapsulated within these spectra.

\begin{figure}[h]
	\centering
	\includegraphics[width=0.7\linewidth]{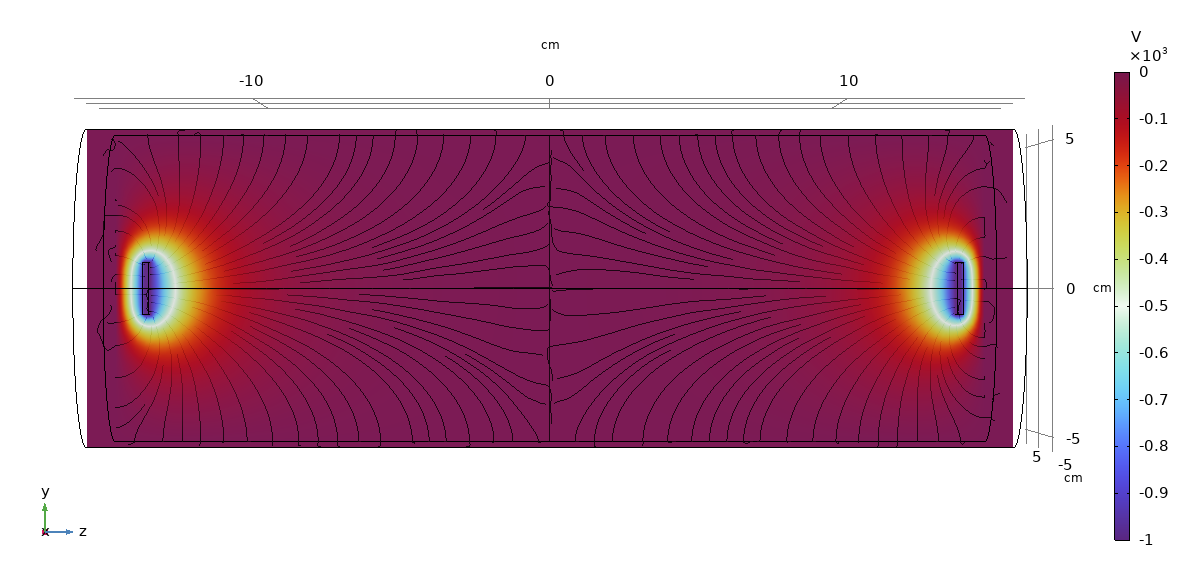}
 \caption{Simulation result of the electrostatic field.}
	\label{fig:pic4}
\end{figure}

To mitigate measurement deviations caused by the cavity structure on both sides of the membrane, we designed the cavity to exhibit a symmetrical structure. We performed simulations of the electrostatic field distribution to validate this design choice under the system's current geometry and material properties. The electric field was analyzed through simulations with \textit{COMSOL Multiphysics} (a simulation platform that offers both fully coupled multiphysics and single-physics modeling capabilities),  and the simulation results are illustrated in Fig.$\ref{fig:pic4}$. As anticipated, the figure displays a well-defined symmetrical electrostatic field structure. This symmetric structure is crucial as it effectively minimizes the impact of asymmetry on particle detection efficiency on both sides. At the same time, the simulation results also demonstrate that the electric field distribution inside the detector chamber is favorable for collecting positively charged radon progeny. Specifically, the radon progeny located at the center of the chamber will be effectively drawn toward the vicinity of the semiconductor surface by the electric field.

\subsection{Characteristic Energy Peaks of Radon Daughters Measurement}

\begin{figure}[ht]
	\centering
\includegraphics[width=0.7\linewidth]{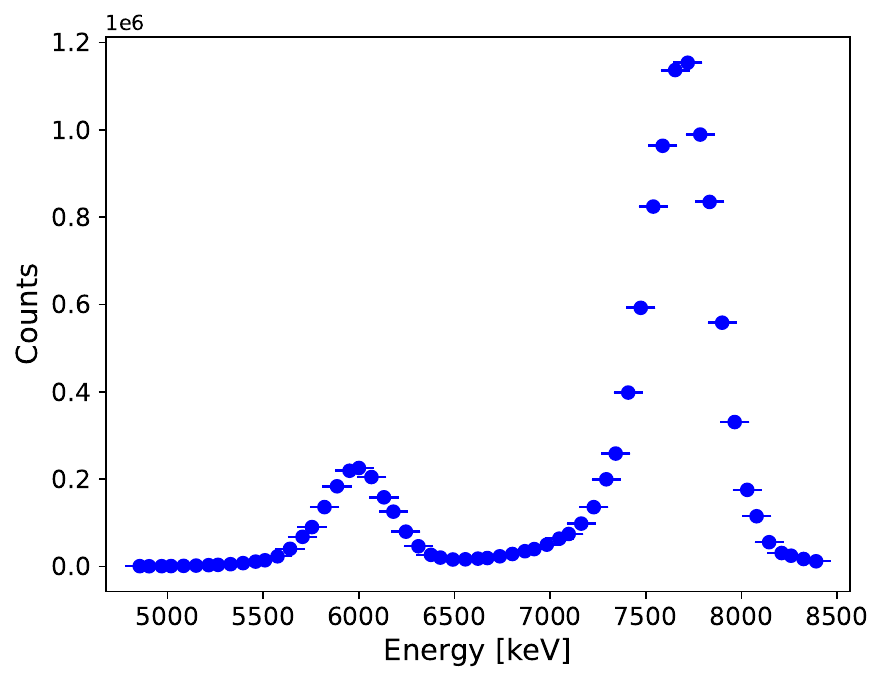}
 \caption{The two $\alpha$ characteristic peaks from left to right correspond to ${}^{218}$Po (6002.4 keV) and ${}^{214}$Po (7686.8 keV).}
	\label{fig:pic5}
\end{figure}

In the decay chain of ${}^{222}$Rn, our detector can distinctly identify several monoenergetic $\alpha$ peaks. As illustrated in Fig.\ref{fig:pic5}, the figure displays two prominent peaks, arranged from left to right, corresponding to the peaks of ${}^{218}$Po (6002.4 keV) and ${}^{214}$Po (7686.8 keV). 

An advantageous aspect of ${}^{214}$Po's position further along the decay chain compared to ${}^{218}$Po is its positive impact on the collection efficiency. The intermediate isotopes, ${}^{214}$Pb and ${}^{214}$Bi, are also pulled to the silicon PIN diode surface by the electric field and decay there, allowing the subsequent ${}^{214}$Po decays to be detected effectively. Consequently, this enhances the overall event collection efficiency of  ${}^{214}$Po. As Fig.\ref{fig:pic5} demonstrates, the characteristic peak of ${}^{214}$Po exhibits a higher event count within the same collection time. Additionally, there is the substantial overlap between the energy peaks of ${}^{210}$Po (5304.3 keV) and ${}^{218}$Po, presenting a significant challenge in distinguishing between these events. Although the rate of these ${}^{210}$Po events is so low that they are nearly invisible in the plot. Considering these two factors, our subsequent analyses consistently prioritize selecting the energy peak of ${}^{214}$Po as the measurement for nuclear radioactivity, thereby enhancing the precision of our measurements.

\subsection{Calibration and Background}

To calibrate the efficiency of our system, we connected the radon source, diffusion chamber, and \href{https://durridge.com/products/rad7-radon-detector/}{RAD7} (a versatile radon and thoron detector from DURRIDGE) in sequence, flushing the entire system with nitrogen throughout the process at a flow rate of 1 SLPM. After turning on the ${}^{226}$Ra source, we waited until the radon concentration reached equilibrium before taking measurements. The concentration of ${}^{214}$Po can be inferred from the measured concentration of ${}^{222}$Rn in RAD7. Through calibration, we obtained the detection efficiency of the system for ${}^{214}$Po as shown in Table \ref{tab:table1}. The calibration efficiency of each side of the detector can be obtained by calculating the ratio of the radon number concentration measured by each detector to the concentration measured by the RAD7.

\begin{table}[ht]
\centering
\caption{Calibration for the detection efficiency}
\label{tab:table1}
\small
\begin{tabular}{ccc}
\hline
 Detection position & ${}^{214}$Po & Detection efficiency\\
 & (Bq/m$^3$) &\\
\hline
RAD7 & 182.0$\pm$4.0  & -  \\
Detector Side A & 55.9$\pm$0.6  & (30.7$\pm$0.8)\% \\
Detector Side B & 49.9$\pm$0.6  & (27.4$\pm$0.7)\% \\
\hline
\end{tabular}
\end{table}

There are two possible explanations for this 3\% efficiency gap on each side. On the one hand, even though we have tried to design the whole system as a symmetrical structure to reduce the difference between the two sides, some factors that are not easy to control, such as the differences introduced by hand during the pre-signal amplifier welding process, the silicon pin placement angle, etc., may cause the acquisition efficiency of the detector at the two ends to be different. However, since we maintain a consistent angle for the two silicon PIN diodes in each trial and consistently employ the same pre-signal amplifier, with a predetermined correspondence between the silicon PIN and the pre-signal amplifier, any efficiency discrepancies arising from these factors can be considered fixed and should not impact the accuracy of our experiments. 

On the other hand, during the calibration process, despite our continuous use of nitrogen flushing to enhance gas flow within the system, the system's geometry may give rise to gas short-circuiting. In other words, the chamber creates a shorter gas pathway between the nitrogen inlet and the outlet. Consequently, when nitrogen is blown in, it exits through this gas pathway at a faster rate compared to its flow through the entire chamber space. That can result in non-uniform radon concentration within the chamber, causing a concentration disparity between the gas pathway and the distant regions of the detector. As a result, the radon concentration measured at the far end may be slightly lower than that at the opposite end. This is also why, for our measurements, we chose to let radon diffuse into the detector rather than be blown in by airflow. This issue can be addressed through subsequent experiments aimed at optimizing the design of the chamber's structure.

\begin{figure}[t]
\begin{center}
   \subfloat[Nylon] {
   \includegraphics[width=0.25\textwidth]{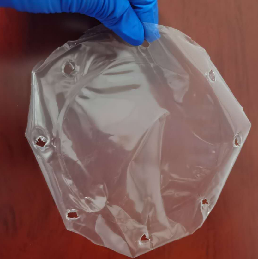}\hfill
   }\quad\quad
   \subfloat[Remoistening masking paper (\href{https://www.daio-paper.co.jp/en/product/paper/}{Daio Paper Corporation})] {
   \includegraphics[width=0.28\textwidth]{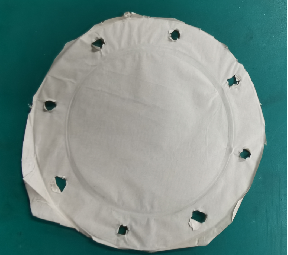}
   }
   
   \subfloat[Light blocking film] {
   \includegraphics[width=0.26\textwidth]{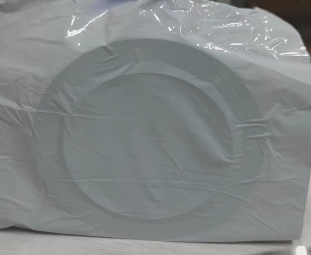}\hfill
   }\quad\quad
   \subfloat[Polyethylene] {
   \includegraphics[width=0.25\textwidth]{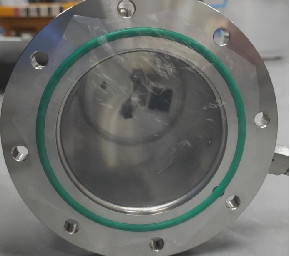}
   }
\end{center}
\caption{The four membrane materials that we tested.}
\label{fig:figurescombine}
\end{figure}

To check the background level of the diffusion chamber, we run the system with the ${}^{226}$Ra source switched off. The signal rate of ${}^{214}$Po on both sides of the membrane is at the level of $0.4$ Bq/m$^3$. Through several comparative experiments, we have observed that the order of magnitude of background events remains consistent across each trial. Compared to the previous subsection, the detector's background level is negligible relative to the radon concentration when the radiation source is open. Additionally, since the background levels vary across experimental measurements, we have chosen to disregard the background influence in subsequent calculations without further adjustment.

\section{Results}

\begin{figure}[!b]
	\centering
\includegraphics[width=0.7\linewidth]{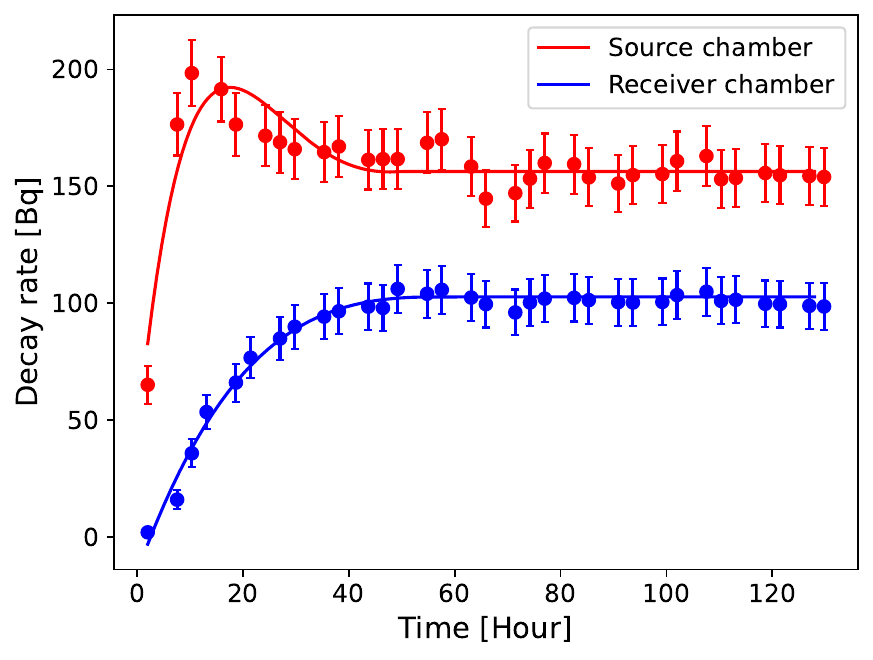}
 \caption{The evolution over time of the radioactive activity of ${}^{214}$Po on both sides of the remoistening adhesive masking paper. The dots in the figure represent experimental data points, while the solid lines indicate the trend fitting curves. Approximately 56 hours after opening the radiation source, the decay rate on both sides gradually stabilizes.}
	\label{evolution}
\end{figure}

Using the experimental setup and methods described in the previous sections, we conducted measurements across several membrane materials currently employed in the Particle and Astrophysical Xenon Experiments (PandaX)\cite{PhysRevLett.127.261802} and JUNO experiments, as shown in Fig.\ref{fig:figurescombine}. Nylon films are often used to protect detector materials with ultra-low background surface treatments during transportation, minimizing radon contamination during transport and assembly. The paper film shown in (b) been used in the JUNO experiment, attached to the acrylic surface during detector construction to prevent contamination from the air radon and dust, and film (c) has beed applied over the paper films to provide additional light and radon shielding. The film (d) is also used in experiments with strict low-background requirements for device and parts packing.

After a specific duration, we observed that the concentration of ${}^{214}$Po in the chamber had stabilized on both sides of the membrane (see Fig.\ref{evolution}), indicating that equilibrium had been established in the radon diffusion process. As anticipated, the radon concentrations on either side of the membrane differ due to the presence of the membrane itself. Based on the principles outlined in Section 3., we can determine the ratio of radon concentrations on both sides of the membrane in the equilibrium state by measuring the concentration of ${}^{214}$Po. This relative ratio serves as a crucial parameter for assessing the membrane's effectiveness in inhibiting radon diffusion.

\begin{table}[ht]
    \centering
    \caption{List of parameters of measured materials (25 ℃)}
    \label{tab:table2}
    \small
        \begin{tabular}{ccc}
            \hline
            & Nylon & Remoistening masking paper\\
            \hline
            Translucent & \ding{51} & \ding{55}\\
            $d$ (mm) & 0.083$\pm$0.002 & 0.053$\pm$0.002 \\
            Feature & Without glue & Single-sided water-soluble adhesive\\
            $\eta_R$ & (0.1$\pm$0.01)\% & (72.4$\pm$2.60)\%\\
            $D_a$ (cm$^2$/s) & (2.28$\pm$0.29)$\times$10$^{-10}$ & (4.40$\pm$0.50)$\times$10$^{-7}$\\
            \hline
        \end{tabular}

    \vspace{0.5cm}

        \begin{tabular}{cccc}
            \hline
            & Light blocking film & Polyethylene\\
            \hline
            Translucent& \ding{55} & \ding{51}\\
$d$ (mm) & 0.077$\pm$0.002 &0.089$\pm$0.002 \\
Feature & Without glue &Single-sided minimal adhesive\\
$\eta_R$ & (92.1$\pm$3.40)\%&  (44.1$\pm$1.60)\%\\
$D_a$ (cm$^2$/s) &(2.80$\pm$1.20)$\times$10$^{-6}$   & (2.21$\pm$0.11)$\times$10$^{-7}$  \\
            \hline
        \end{tabular}
\end{table}

Table \ref{tab:table2} shows the corresponding parameters of several measured membrane materials. It is important to note that all measurements have been conducted at room temperature (25 ℃). Here, $d$ is the average thickness of the membrane materials and $\eta_R$ denotes the relative radon ratio of each side, taking into account detection efficiency. Specifically, given that $C_A$ ($C_B$) is the radon concentration on side $A$ ($B$), $E_i$ is the calibration efficiency of side $i$, the calculation expression of $\eta_R$ is as follows:
\begin{equation}
    \eta_R=\frac{C_A/E_A}{C_B/E_B}
\end{equation}

It is evident from the results that different materials offer a significant range of diffusion coefficients, spanning approximately four orders of magnitude. This wide variation highlights the importance of selecting appropriate materials in applications where radon diffusion needs to be minimized. 

\begin{figure}[!h]
	\centering
\includegraphics[width=0.7\linewidth]{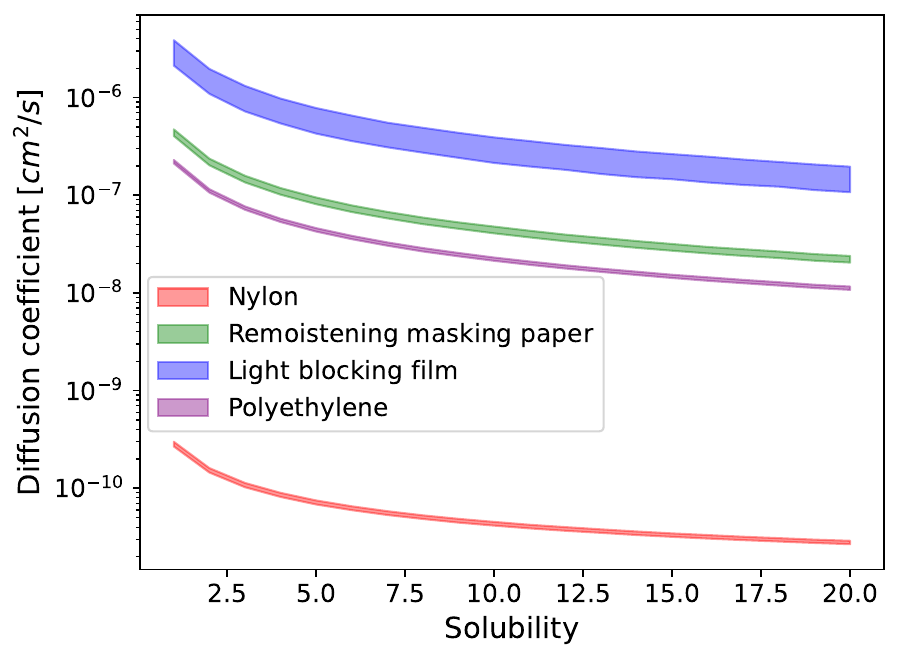}
 \caption{The measurement results of the relative relationship between diffusion coefficient and solubility. The curve band in the figure is the result obtained by scanning $S$ from 1-20 and bringing back the equation (\ref{eq7}). }
	\label{fig:pic6}
\end{figure}

Regarding the relationship between the diffusion coefficient and solubility, as discussed in Section 2., we have to scan the value for solubility $S$ from 1 to 20 and bring it into the equation (\ref{eq8}) to solve the corresponding diffusion coefficient $D$. Still, the negative correlation between the two can be observed from the curve in Fig.\ref{fig:pic6}. The curves for each material represent their respective diffusion coefficients across different solubility values, with the band width indicating the uncertainty at a given solubility. Nylon shows a very low diffusion coefficient, indicating that it effectively impedes the radon diffusion. The light-blocking film has the highest diffusion coefficient, around $10^{-6}$ cm$^2$/s, showing it is less effective at blocking diffusion compared to the other materials. 

\begin{figure}[ht]
	\centering
\includegraphics[width=0.7\linewidth]{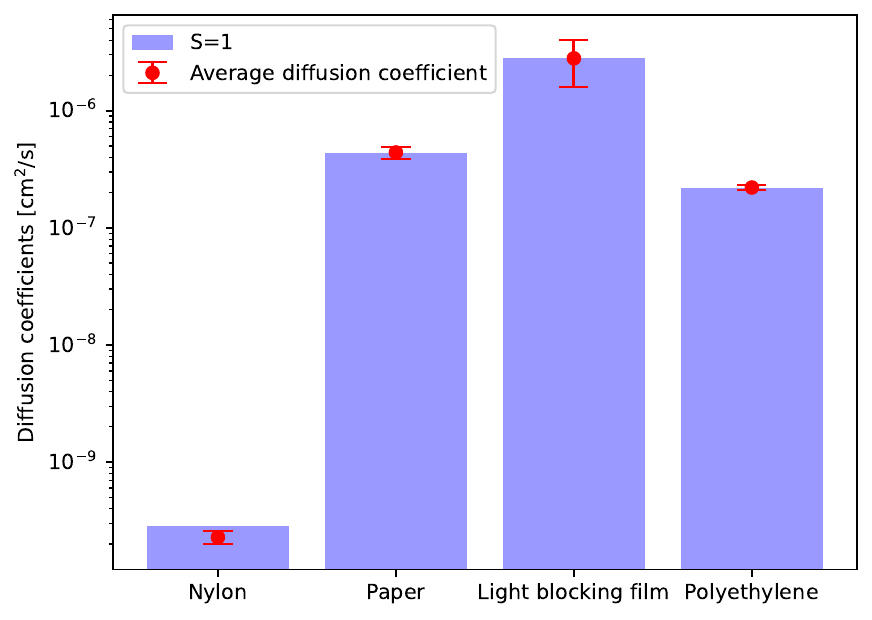}
 \caption{The comparison between average diffusion coefficients for different materials. The red dots represent the average diffusion coefficient and error for different materials. In contrast, the height of the blue shading represents the diffusion coefficient obtained for the material under the assumption of $S=1$.}
	\label{fig:pic7}
\end{figure}

What is worth noting is that the average diffusion coefficients of different materials closely resemble the diffusion coefficients when $S=1$ in the corresponding material curves. Fig.\ref{fig:pic7} compares the average diffusion coefficient values obtained from the experiments and the theoretical values of the diffusion coefficient solved at $S=1$. We can intuitively observe that these two values are almost identical for all materials. This conclusion also provides strong support for our use of an easily calculable average diffusion coefficient, which is calculated from equation (\ref{eq5}). This coefficient relies solely on the difference in radon number concentration across the membrane, enabling a straightforward estimation of the order of magnitude of the material's radon diffusion coefficient. Moreover, this figure can also clearly compare the average diffusion coefficients of different materials.

\section{Conclusion}

We have introduced a symmetric radon diffusion chamber design that builds upon traditional methods and offers enhanced capabilities to measure the diffusion coefficient of radon across membrane materials. Materials like Nylon, which demonstrated a low diffusion coefficient, could be particularly beneficial for experiments demanding stringent controls over radon intrusion. Our results also underscore the influence of solubility on radon diffusion, and our findings on the impact of solubility have broader implications. Rare decay experiments that utilize membrane materials need to consider not only the inherent properties of the material but also factors that could influence its solubility, such as ambient temperature and humidity. Understanding the relationship between the diffusion coefficient and solubility for different materials is crucial for evaluating the ability of materials to block radon diffusion.

\section*{Acknowledgments}

This work has been supported by the National Natural Science Foundation of China (No. 12222505), the Ministry of Science and Technology of China (grant No. 2023YFA1606200) and Shanghai Pilot Program for Basic Research — Shanghai Jiao Tong University (No. 21TQ1400218).



\section*{References}

\bibliographystyle{elsarticle-num}
\bibliography{references}

\end{document}